\documentclass[aps,pre,twocolumn,superscriptaddress,amsmath,showpacs,amssymb]{revtex4}
\usepackage{psfig}
\usepackage{graphics}
\usepackage{graphicx}
\usepackage{epsfig}
\usepackage{psfig}
\usepackage{amsmath}

\def\bc{\begin{center}}
\def\ec{\end{center}}
\def\be{\begin{equation}}
\def\ee{\end{equation}}
\def\la{\langle}
\def\ra{\rangle}

\def\rarr{\rightarrow}
\def\bear{\begin{eqnarray}}
\def\eear{\end{eqnarray}}

\begin{document}

\title{Excitation lines and the breakdown of Stokes--Einstein
relations in supercooled liquids}

\author{YounJoon Jung}

\affiliation{Department of Chemistry, University of California,
Berkeley, CA 94720-1460}

\author{Juan P. Garrahan} 

\affiliation{Theoretical Physics, University of Oxford, 1 Keble Road,
Oxford, OX1 3NP, UK}

\affiliation{School of Physics and Astronomy, University of
Nottingham, Nottingham, NG7 2RD, UK}

\author{David Chandler}

\affiliation{Department of Chemistry, University of California,
Berkeley, CA 94720-1460}

\date{\today}

\begin{abstract}
By applying the concept of dynamical facilitation and analyzing the
excitation lines that result from this facilitation, we investigate
the origin of decoupling of transport coefficients in supercooled
liquids.  We illustrate our approach with two classes of models.  One
depicts diffusion in a strong glass former, and the other in a fragile
glass former.  At low temperatures, both models exhibit violation of
the Stokes-Einstein relation, $D\sim\tau^{-1}$, where $D$ is the self
diffusion constant and $\tau$ is the structural relaxation time.  In
the strong case, the violation is sensitive to dimensionality $d$,
going as $D\sim\tau^{-2/3}$ for $d=1$, and as $D\sim \tau^{-0.95}$ for
$d=3$.  In the fragile case, however, we argue that dimensionality
dependence is weak, and show that for $d=1$, $D \sim \tau^{-0.73}$.
This scaling for the fragile case compares favorably with the results
of a recent
experimental study for a three-dimensional fragile glass former.
\end{abstract}

\maketitle

\section{Introduction}
Normal liquids exhibit homogeneous behavior in their dynamical
properties over length scales larger than the correlation length of
density fluctuations. For example, the Stokes--Einstein relation
that relates the self--diffusion constant $D$, viscosity $\eta$, and
temperature $T$,
\be D \propto {T \over \eta}, \label{SE} \ee 
is usually accurate \cite{hansen,balucanizoppi}.  This relation is
essentially a mean field result for the effects of a viscous
environment on a tagged particle. In recent experimental studies, it
has been reported that the Stokes--Einstein relation breaks down as
the glass transition is approached in supercooled liquid systems
\cite{ediger-arpc-00,fujara-zphyb-92,chang-jpcb-97,cicerone-jcp-96,blackburn-jpc-96,swallen-prl-03}.
Translational diffusion shows an enhancement by orders of magnitude
from what would be expected from Eq.\ (\ref{SE})
\cite{hodgdon-pre-93,stillinger-pre-94,tarjus-jcp-95,liu-pre-96,xia-jpcb-01}. Here,
we show that this breakdown is due to fluctuation dominance in the
dynamics of low temperature glass formers.  These pertinent
fluctuations are dynamic heterogeneities
\cite{perera-pre-96,perera-jcp-99,perera-jncs-98,kob-prl-97,donati-prl-98,donati-pre-99,donati-prl-99,glotzer-jncs-00}.
Thus, the Stokes--Einstein breakdown is one further example of the
intrinsic role of dynamic heterogeneity in structural glass formers
\cite{palmer-prl-84,garrahan-prl-02,garrahan-pnas-03}.

In the treatment we apply, dynamic heterogeneity is a manifestation of
excitation lines in space--time \cite{garrahan-prl-02}.  This picture
leads to the prediction of dynamic scaling in supercooled liquids,
$\tau(l) \sim l^{z}$.  Here, $\tau(l)$ is the structural relaxation
time for processes occurring at length scale $l$, and $z$ is a dynamic
exponent for which specific results have been established
\cite{garrahan-prl-02,garrahan-pnas-03,whitelam-condmat-03}.  This
picture and its predicted scaling results differ markedly from those
derived with the view that glass formation is a static or
thermodynamic phenomenon
\cite{debenedetti-nature-01,sastry-nature-98,buchner-pre-99,buchner-prl-00,schroder-jcp-00,keyes-pre-02,kivelson-physa-95,xia-pnas-00}.
It also differs from mode coupling theory which predicts singular
behavior at non--zero temperature \cite{gotze-repprog-92,kob-rev-03}.

This paper is organized as follows.  In Sec.~\ref{sec2}, we introduce
our model for a supercooled liquid with a probe molecule immersed in
the liquid.  Simulation results are given in Secs.~\ref{sec3} and
\ref{sec4}. Section \ref{sec4} also provides analytical analysis of
the diffusion coefficient and the Stokes--Einstein violation, and
explains the origin of the decoupling of transport coefficients based
on the excitation line picture of trajectory space. Comparison of our
theory with recent experimental results is carried out in
Sec.~\ref{sec5}. We conclude in Sec.~\ref{sec6} with a Discussion.

\section{Models} \label{sec2}
We imagine coarse graining a real molecular liquid over a microscopic
time scale (e.g., larger than the molecular vibrational time scale),
and also over a microscopic length scale (e.g., larger than the
equilibrium correlation length). In its simplest form, we assume this
coarse graining leads to a kinetically constrained model
\cite{garrahan-prl-02,garrahan-pnas-03,fredrickson-prl-84,jackle-zphysb-91,ritort-advphys-03}
with the dimensionless Hamiltonian,
\be H=\sum_{i=1}^{N} n_{i}, \ \ (n_i=0, 1). \label{hamil} \ee
Here, $n_i=1$ coincides with lattice site $i$ being a spatially
unjammed region, while $n_{i}=0$ coincides with it being a jammed
region.  We call $n_i$ the ``mobility field''.  The number of sites,
$N$, specifies the size of the system.  From Eq.~(\ref{hamil}),
thermodynamics is trivial, and the equilibrium concentration of
defects or excitations is
\be c=\langle n_i\rangle = { 1\over 1 + \exp ({1/\widetilde{T}})}, \ee
where $\widetilde{T}$ is a reduced temperature.  We make 
explicit connection of $\widetilde{T}$ with absolute temperature later
when comparing our theory with experimental results.

The dynamics of these models obey detailed balance and local dynamical
rules. Namely,
\begin{equation}
n_i=0 
\begin{array}{c}
\xrightarrow{k_{i}^{(+)}} \\
\xleftarrow[k_{i}^{(-)}]{} \\
\end{array}
n_i=1 ,
\end{equation}
where the rate constants for site $i$, $k_{i}^{(+)}$ and
$k_{i}^{(-)}$, depend on the configurations of nearest neighbors.  For
example, in dimension $d=1$,
\begin{eqnarray}
k_{i}^{(+)} &=& e^{-1/\widetilde{T}} f(n_{i-1},  n_{i+1}), \\
k_{i}^{(-)} &=& f(n_{i-1},  n_{i+1}),
\end{eqnarray}
where $f(n_{i-1},n_{i+1})$ reflects the type of dynamical
facilitation.  In the Fredrickson--Andersen (FA) model
\cite{fredrickson-prl-84}, a state change is allowed when it is next
to at least one defect.  The facilitation function in this case is
given by,
\begin{equation}
 f_{\rm FA}(n_{i-1},  n_{i+1}) = n_{i-1} + n_{i+1} - n_{i-1} n_{i+1} .
\end{equation}
In the East model \cite{jackle-zphysb-91},
dynamical facilitation has directional persistence. The 
facilitation function in this case is
\begin{equation}
f_{\rm East}(n_{i-1},  n_{i+1}) = n_{i-1} .
\end{equation}

In order to study translational diffusion in supercooled liquids, we
extend the concept of dynamic facilitation to include a probe
molecule. The dynamics of a probe will depend on the local state of
the background liquid. When and where there is no mobility, the
diffusive motion of the probe will be hindered. When and where there
is mobility, the probe molecule will undergo diffusion easily. As
such, in a coarse--grained picture, the probe molecule is allowed to
jump from lattice site $i$ to a nearest neighbor site when site $i$
coincides with a mobile region, $n_i=1$. In order to satisfy detailed
balance, we further assume that the probe molecule can move only to a
mobile region, i.e.,
\be x(t+\delta t) = x(t) \pm \delta x \cdot n_x \cdot n_{x\pm \delta x}, \ee
where $x(t)$ denotes the position of the probe at time $t$. Units of
time and length scales are set equal to a Monte Carlo sweep and a
lattice spacing, respectively.

\section{Computer Simulations}
\label{sec3}

Using the rules described in Sec.~\ref{sec2}, we have performed Monte
Carlo simulations of diffusion of a probe molecule in the FA and East
models for various temperatures.  For the purpose of numerical
efficiency, we have used the continuous time Monte Carlo algorithm
\cite{bkl-jcompphys-75,newmanbarkema}. In the all systems, $N$ was
chosen as $N=100/c$, and the simulations were performed for total
times ${\cal T}\approx 100\tau$, with $\tau$ being the relaxation time
of the model. Averages were performed over $10^3$ to $10^5$
independent trajectories.

\begin{figure}[ht]
\begin{center}
\epsfig{file=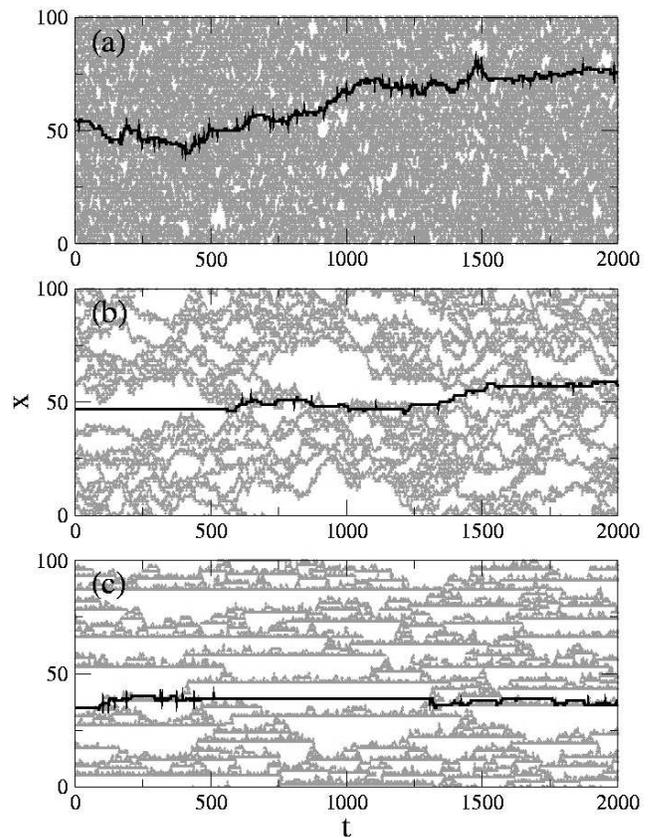, width=8.4cm}
\caption{Typical trajectories of a probe molecule in one--dimensional
models. The probe molecule (black line) undergoes a diffusive process
in the trajectory space that consists of gray (mobile) and white
(immobile) regions. (a) FA model at $\widetilde{T}=3$; (b) FA model at
$\widetilde{T}=0.8$, and (c) East model at
$\widetilde{T}=0.8$. \label{fig1}}
\end{center}
\end{figure}

In Fig.~\ref{fig1}, we show typical trajectories of probe molecules in
the FA and East models. In the high temperature case, trajectory space
is dense with mobile regions and there are no significant patterns in
space--time. As such, the dynamics is mean--field like. It is for this
reason that the relaxation time in this case is inversely proportional
to the equilibrium probability of excitation, $c$ (see, for example,
Ref.~\cite{berthier-pre-03}).  The probe molecule executes diffusive
motion, without being trapped in immobile regions for any significant
period of time.

The low temperature dynamics is different.  Mobility is sparse,
defects tend to be spatially isolated at a given time, and trajectory
space exhibits space--time patterns. See Fig.~\ref{fig1}(b) and (c).
Because of the facilitation constraint, an immobile region needs a
nearest mobile region to become mobile at a later time. The
excitations therefore form continuous lines and bubble--like
structures in trajectory space. While inside a bubble, the probe
molecule will be immobilized.  See, for example, the segment of the
trajectory of a probe molecule for $0<t<500$ in Fig.~\ref{fig1}(b).
Due to exchanges between mobile and immobile regions, an immobile
region can become mobile after a period of time. At that stage the
probe molecule can perform a random walk until it is again in an
immobile region. The motion of a probe molecule will manifest
diffusive behavior over a time long enough for many dynamical
exchanges to occur. In the East model at low temperatures such as
pictured in Fig.~\ref{fig1}(c), the bubbles in space--time form
hierarchical structures \cite{garrahan-prl-02}.

\begin{figure}[t]
\begin{center}
\epsfig{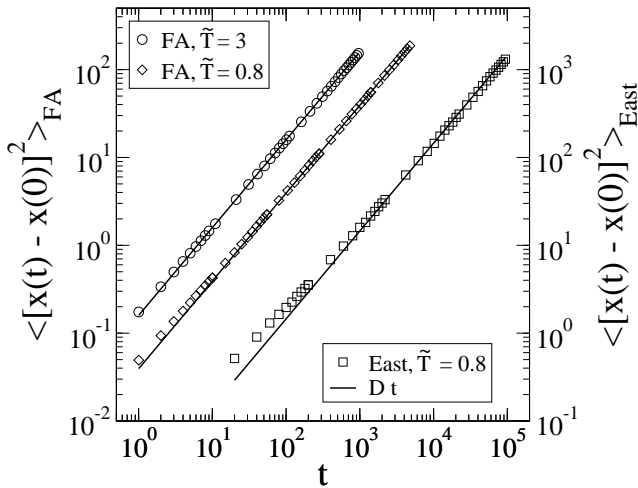}
\caption{Mean squared displacements of the probe molecules are shown
for the three different cases illustraed in Fig.~\ref{fig1}.
\label{fig2}}
\end{center}
\end{figure}

Figure~\ref{fig2} plots mean square displacements of probe molecules
for the FA and East models for three different cases pictured in
Fig.~\ref{fig1}. In the high temperature case, the mean square
displacement reaches its diffusive linear regime after a very short
transient time. In the low temperature case, the probe molecule in the
East model case reaches the diffusive regime after a longer time and
over a larger length scale than that in the FA model with the same
reduced temperature.

\section{Stokes--Einstein Violation}
\label{sec4}

\subsection{Diffusion Coefficient}

\begin{figure}[t]
\begin{center}
\epsfig{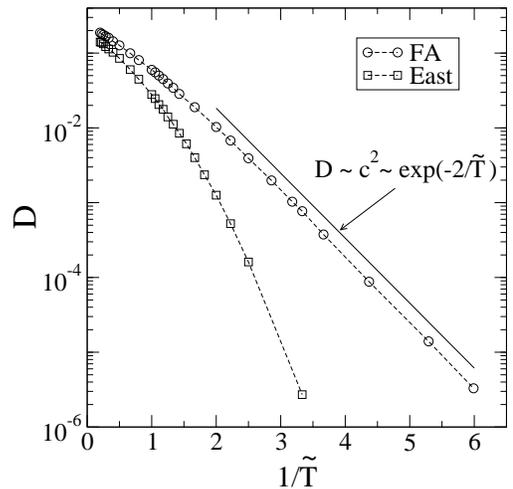}
\caption{Diffusion coefficients for the FA and East models as
functions of $1/\widetilde{T}$.  Dashed lines are a guide to the
eye. \label{fig3}}
\end{center}
\end{figure}

Figure~\ref{fig3} plots the diffusion coefficient of a probe molecule
for the FA and the East models. The diffusion coefficient is
determined from the mean square displacement,
\be D=\lim_{t\rarr\infty} {\la [\Delta x(t)]^2 \ra \over t}, \ee
where $\Delta x(t) = x(t)-x(0)$.  Error estimates for our simulations
are no larger than the size of the symbols.

In the FA model, the diffusion coefficient exhibits Arrhenius behavior
for $\widetilde{T}<1$.  This behavior reflects the fact that
relaxation dynamics in the FA model is similar to that of a strong
liquid.  In this regime, over more than 4 orders of magnitude in $D$,
the slope of $\log D$ vs $\widetilde{T}^{-1}$ is close to 2.  This
result is consistent with the expected low temperature scaling,
\be D_{\rm FA} \sim c^2 \sim \exp (-2/\widetilde{T}), \label{Dfa} \ee
as discussed in the next subsection.  In the East model case, also
pictured in Fig.~\ref{fig3}, the diffusion coefficient decreases more
quickly than Arrhenius. This super--Arrhenius behavior is due to the
hierarchical nature of dynamics in the East model
\cite{sollich-prl-99}.

\begin{figure}[t]
\begin{center}
\epsfig{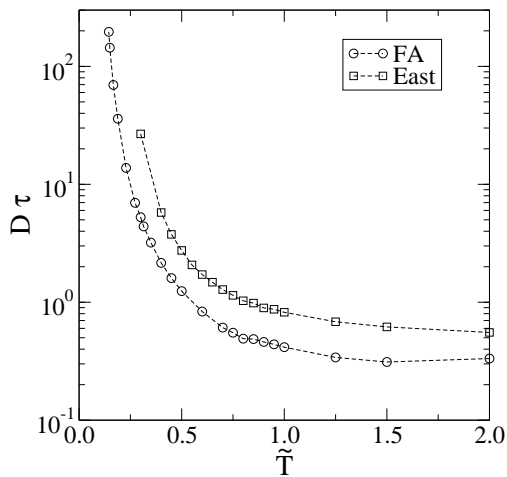}
\caption{Violations of the Stokes--Einstein relation are similar in
  the $d=1$ FA and East models. Dashed lines are a guide to the
  eye.\label{fig4}}
\end{center}
\end{figure}

Comparing the diffusion coefficients with the relaxation times of the
background liquids demonstrates Stokes--Einstein violation in both
models. The relaxation times, $\tau$, of the FA and the East models at
different temperatures have been determined in prior work
\cite{buhot-pre-01,berthier-jcp-03}. When the Stokes--Einstein
relation is satisfied, $D\tau\sim {\rm const}$.  This behavior occurs
in the FA and East models when $\tilde{T}>2$, but Fig.~\ref{fig4}
shows that $D\tau$ is enhanced from that behavior by 2 or 3 orders of
magnitude when $\tilde{T}<1$.  Bear in mind, these deviations from
Stokes--Einstein are $d=1$ results. The appropriate generalization of
the FA model to $d=3$ does not exhibit such large deviations. On the
other hand, we expect that generalizations of the East model, which is
hierarchical and therefore fragile, will have weak dimensional
dependence and continue to exhibit large deviations for $d=3$.  We
turn to the arguments that explain these claims now.

\subsection{Scaling Analysis}

For high temperatures, the local mobility field will tend to be close
to its mean value, $c$.  As such, both the relaxation mechanism of the
material and the diffusional motion of the probe molecule make use of
the same local mobility fields. For this reason, the diffusion
coefficient and the relaxation time scale are strongly coupled in this
regime, leading to the Stokes--Einstein relation.

At low temperatures, however, the dynamics of the system is
not so simply related to the mean mobility field. Here, the
fluctuations of bubble structures dominate. The relaxation time of the
background liquid will approximately scale as the longest temporal
extension of bubbles. The {\it persistence time} of an individual
lattice site, $t_{\rm pers}$, is the time for which that site makes
its first change in state.  Its typical size will be intimately tied
to the structural relaxation time of the liquid. For the FA model in
$d=1$,
\be \tau \sim \langle t_{\rm pers} \rangle \sim c^{-3}. \label{taufa} \ee
See, for example, Refs.~\cite{ritort-advphys-03,garrahan-prl-02}.

This result is consistent with a simple argument concerning diffusive
motions of excitation lines in the low temperature FA model
\cite{garrahan-prl-02}. In particular, the structural relaxation times
in the FA model is given by the time in which a typical bubble
structure looses its identity through wandering motions of excitation
lines. The excitation line has a local diffusivity of ${\cal D}\sim
c$. (We use caligraphic ${\cal D}$ to distinguish this diffusion
constant for excitations from that for particles, $D$.)
In order to form a bubble, an excitation line needs to wander distance of the
order of the typical length between defects, $l_{\rm eq}\sim
c^{-1}$. Therefore, the mean relaxation time is given by $\tau\sim
l_{\rm eq}^2/{\cal D}\sim c^{-3}$.

\begin{figure}[t]
\begin{center}
\epsfig{file=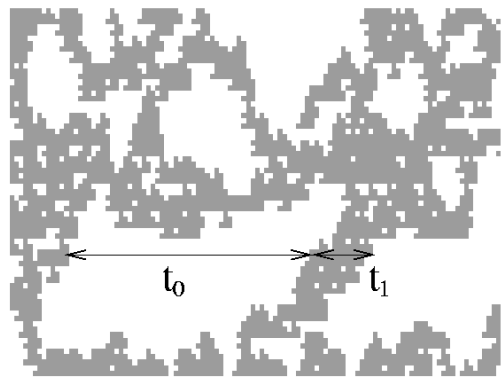, width=6.6cm}
\caption{A section of Fig.~\ref{fig1}(b) illustrating the meanings of
exchange times, $t_0$ and $t_1$. $t_0$ is a time a site spends in a
bubble, and $t_1$ is a time it spends in a surrounding boundary.
\label{fig5}}
\end{center}
\end{figure}

When the probe molecule is at the boundary of a bubble, it may not
need to wait until the bubble closes in order to undergo diffusion;
rather, it can remain within mobile cells and diffuse around the
boundary of the bubbles. In this way, translational diffusion will be
more facilitated than structural relaxation, leading to an enhanced
diffusion in the fluctuation dominated low temperature
region. Specifically, consider the dynamical {\it exchange times},
i.e., the times between flipping events for a given lattice site. See
Fig.~\ref{fig5}. $t_{0}$ is such a time duration for an $n_{i}=0$
state and $t_{1}$ is such a time duration for an $n_{i}=1$ state. The
probe molecule can move only while in a mobile region. Further, the
mean square displacement of the probe will be proportional to the
number of diffusive steps that a probe molecule will take during the
trajectory, ${\cal N}$,
\be \la [\Delta x(t)]^2\ra \sim {\cal{N}} \sim {{\cal T}\over \la t_{0}\ra + \la t_{1}\ra}.
\label{msqd-FA} \ee
Here, $\cal{T}$ is the length of a long trajectory in the FA
model. The average duration of the defect state, $\la t_{1}\ra$, is
inversely proportional to the probability of a lattice site being
mobile, $c$, times the flip rate, $k_{i}^{(-)}$. Since $k_{i}^{(-)}
\sim {\cal O}(1)$, we have,
\be \la t_{1} \ra \sim c^{-1}. \label{t1} \ee
From detailed balance, therefore,
\be \la t_{0} \ra \sim c^{-2}. \label{t0} \ee
Since $\la t_{1}\ra \ll \la t_{0}\ra$ in the low temperature region,
Eqs.~(\ref{msqd-FA})--(\ref{t0}) give
\be D_{\rm FA} \sim {\la (\Delta x)^2\ra \over {\cal T}} \sim {1\over
\la t_{0}\ra} \sim c^{2}. \ee
This result explains Eq.~(\ref{Dfa}). Together with Eq.~(\ref{taufa}),
it leads to
\be D_{\rm FA} \sim \tau^{-\xi}, \ee
with $\xi = 2/3$ in the $d=1$ FA model case. This scaling is to
be contrasted with the Stokes--Einstein result, $\xi=1$.

Numerical simulation \cite{garrahan-pnas-03} and renormalization group
analysis \cite{whitelam-condmat-03} of higher dimension
generalizations of the FA model indicate that for $d=3$, $\tau\sim
c^{-2.1}$. However, the scaling $D\sim c^2$ remains true for all
dimensions as it is based solely on detailed balance. Thus, for $d=3$,
$\xi\approx 0.95$. In other words, there is only a weak breakdown in
the Stokes--Einstein relation for strong liquids in $d=3$.

\begin{figure}[t]
\begin{center}
\epsfig{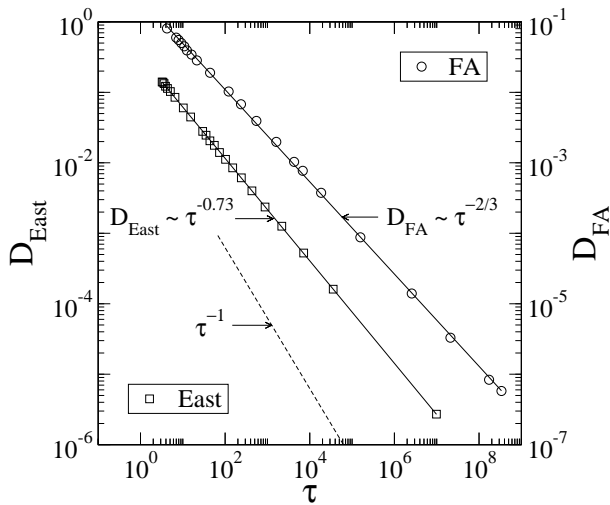}
\caption{Scaling of Stokes--Einstein violation in $d=1$. 
Circles and squares indicate computed results for the FA and East
models, respectively.  \label{fig6}}
\end{center}
\end{figure}

In the East model case, both the diffusion coefficient and the
relaxation time show super-Arrhenius behavior.  The hierarchical,
fractal structure of pattern development in trajectory space for the
East model does not allow a simple scaling analysis of the diffusion
coefficient, and it is not obvious whether temperature independent
scaling exists.  One can define a temperature dependent scaling
exponents, $\alpha(\widetilde{T})$ and $z(\widetilde{T})$,
\begin{eqnarray}
 D_{\rm East} &\sim& c^{\alpha(\widetilde{T})}, \\
\tau &\sim& l^{z(\widetilde{T})},
\end{eqnarray}
so that
\be D_{\rm East} \sim \tau^{-{\alpha(\widetilde{T})/z(\widetilde{T})}}.  \ee
Interestingly, our numerical results indicate that $\xi = {\alpha/z}
\approx 0.73 $ is independent of temperature as shown in
Fig.~\ref{fig6}. This exponent, $\xi \approx 0.73$ for the $d=1$ East
model, is very close to what many experiments and simulations have
found for three--dimensional glass forming liquids. For example, a
recent experiment finds that $\xi \approx 0.77$ in the self--diffusion
of {\it tris}-naphthylbenzene(TNB) \cite{swallen-prl-03}. It was found
that $\xi\approx 0.75$ in a molecular dynamics simulation of
Lennard--Jones binary mixture \cite{yamamoto-prl-98} and a recent
detailed scaling analysis of numerical results shows $\xi \approx
0.65$ \cite{berthier-condmat-03}.

Presumably, such good agreement of scaling relation between the $d=1$
East model and higher dimension systems arises due to directional
persistence of facilitation in the fragile liquid
\cite{garrahan-prl-02,garrahan-pnas-03}.  This persistence in higher
dimensions causes motion to be effectively one--dimensional
\cite{garrahan-pnas-03}. Therefore, dimensionality is not very
significant for fragile glass formers. As such, for fragile systems,
we expect that the scaling relation of the Stokes--Einstein violation
will be reasonably well described by the $d=1$ East model. Based on
this expectation, we further purse the comparison between theory and
experiment.

\section{Comparison with Experiment}
\label{sec5}
Swallen {\it et al.~}\cite{swallen-prl-03} measured the
self--translational diffusion coefficient of TNB near the glass
transition temperature. They observed an increase of $D\eta/T$ from
its high temperature limit by a factor of 400 near the glass
transition temperature.  In order to compare our results with these
experiments, we need to determine the excitation concentration, $c$,
as function of temperature. Since TNB behaves as a fragile liquid, we
determine the excitation concentration as a function of temperature by
fitting the viscosity data of TNB \cite{plazek-TNB} with the
generalization of the East model formulas to higher dimensions
\cite{garrahan-pnas-03}. Namely,
\be \ln{\tau}\approx {1\over d \ln2} \left[\ln(g/c)\right]^2, \label{lntau} \ee
where $g$ is the number of equally likely persistence directions on a
cubic lattice, and
\be \ln(c)= \ln(c_{R})-J\left({1\over T}-{1\over T_{R}}\right). \label{lnc} \ee
The parameter $J$ is the energy scale associated with creating a
mobile region from an immobile region, and $T_{R}$ is an appropriate
reference temperature. Details on the fitting can be found in
Ref.~\cite{garrahan-pnas-03}. Taking $g = 8$ (the cubic lattice value)
and $T_{R}$ as the
temperature at which $\log\tau$ is half the value of
$\log\tau(T_{g})$, we determine that $J/T_{g}\approx 21.7$, and
$\log_{10}(c_{R}/g)\approx -1.28$. The reduced temperature,
$\widetilde{T}$ of the East model is related to absolute temperature
by
\be {1\over \widetilde{T}} = J\left({1\over T}-{1\over T_{R}}\right) +
\ln (g/c_{R}). \ee

Once we have determined the excitation concentration as a function of
the temperature, we can compare experimental data with our computed
results for the Stokes--Einstein violation in the East model
case. Based on the argument that the scaling relation of the
Stokes--Einstein violation ($D\sim \tau^{-\xi}$) remains robust in
higher dimensions and from the dimensional dependence of
Eq.~(\ref{lntau}), we expect
\be \ln{(D\tau)|_{d=3}} \approx {1\over 3}
\ln{(D\tau)|_{d=1}}. \label{East3d} \ee

\begin{figure}[t]
\begin{center}
\epsfig{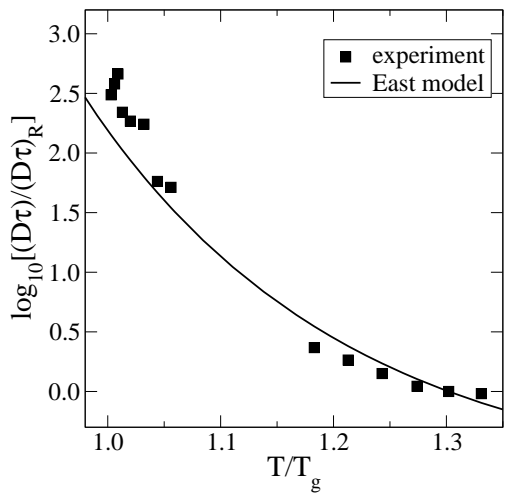}
\caption{Comparison between the East model prediction and experiments
on supercooled TNB, Ref.~\cite{swallen-prl-03}. \label{fig7}}
\end{center}
\end{figure}

In Fig.~\ref{fig7} we use this relationship to compare the extent of
the Stokes--Einstein violation of the experimental system with our
East model results. The agreement between the two appears excellent.

\section{Discussion}
\label{sec6}

There have been previous theoretical studies on the violation of the
Stokes--Einstein relation in supercooled liquid systems. For example,
Kivelson and Tarjus have argued that the Stokes--Einstein violation
can be understood from their ``frustration--limited domain'' model for
supercooled liquids \cite{tarjus-jcp-95,kivelson-physa-95}.  Assuming
a distribution of local relaxation times associated with domain
structures, this model describes the translational diffusion and
viscosity as corresponding to different averaging process of such a
distribution. Their idea contrasts to ours in that the domain
structure in their work is purely static, and the exchange between
different domains are not considered.

Hodgdon and Stillinger have proposed a fluidized domain model
\cite{hodgdon-pre-93,stillinger-pre-94}. In their work, it is assumed
that the system consists of a sparse collection of fluid--like domains
in a background of more viscous media, and fluid--like domains appear
and disappear with a finite life--time and rate. Relaxation times are
determined by the rate of appearance of the fluid--like domains, while
translation diffusion also depends on the life--time of the domains.
To the extent that these domains refer to space--time and not simply
space, this picture is not inconsistent with ours.  Xia and Wolynes
have applied the so--called ``random first order transition theory''
\cite{xia-pnas-00} to the Hodgdon--Stillinger model
\cite{xia-jpcb-01}. In this case, the picture is both mean field and
static and decidedly contrary to our fluctuation dominated and dynamic
view. 

From the perspective that Stokes--Einstein violation is a
manifestation of fluctuation dominated dynamics, one expects that
similar decoupling behavior occurs between other kinds of transport
properties near the glass transition. The extent to which such
decoupling can appear depends upon microscopic details in the specific
transport properties and materials under study. For example, molecular
rotations of a probe will be coupled to the mobility field, but less
so than translations. Indeed, single molecule experiments indicate
that rotations persist in both mobile and immobile regions of a glass
former \cite{deschenes-sci-01,deschenes-jpcb-02,bratko-prl-02}.
Rotational motions can therefore average the effects of dynamic
heterogeneity to a greater extent than translational motions. As such,
decoupling of rotational relaxation from structural relaxation can be
more difficult to detect than violations of the Stokes--Einstein
relation. Precisely how such effects might be detected
seems worthy of further theoretical analysis.

\acknowledgments 

We are grateful to M.D. Ediger, D.A. VandenBout, L. Berthier and
S. Whitelam for discussions.  This work was supported at Berkeley by
the Miller Research Fellowship (YJ)  and by the US Department
of Energy Grant No.\ DE-FG03-87ER13793 (DC), at Oxford by EPSRC Grant No.\
GR/R83712/01 and the Glasstone Fund (JPG), and at Nottingham by EPSRC Grant
No.\ GR/S54074/01 (JPG).

\end{document}